\documentclass{article}
\pdfoutput=1
\usepackage{jheppub}

\usepackage{amsmath}
\usepackage{ulem}

\newcommand{\roughly}[1]{\mathrel{\raise.3ex\hbox{$#1$\kern-0.85em
\lower1ex\hbox{$\sim$}}}}

\newcommand{\gsim}{\roughly>}

\def\exd{{\hbox{d}}}

\def\ba{\begin{eqnarray}}
\def\ea{\end{eqnarray}}
\def\be{\begin{equation}}
\def\ee{\end{equation}}

\def\bfp{{\bf p}}

\def\bfx{{\bf x}}
\def\bfy{{\bf y}}

\def\ssB{{\scriptscriptstyle B}}

\def\ssH{{\scriptscriptstyle H}}
\def\ssM{{\scriptscriptstyle M}}
\def\ssN{{\scriptscriptstyle N}}

\def\ssT{{\scriptscriptstyle T}}

\def\cG{\mathcal{G}}

\def\cO{\mathcal{O}}

\def\nn{\nonumber}

\def\({\left(}
\def\){\right)}

\def\pref#1{(\ref{#1})}

\usepackage{graphicx}
\usepackage{dcolumn}
\usepackage{bm}
\usepackage{color}
\usepackage{ulem}

\title{Failure of Perturbation Theory Near Horizons:\\ the Rindler Example}

\author[a,b]{C.P.~Burgess,}
\author[a,b]{Joshua Hainge,}
\author[a,b]{Greg Kaplanek}
\author[a,b]{and Markus Rummel}
\affiliation[a]{Department of Physics \& Astronomy, McMaster University, Hamilton, Ontario, L8S 4M1, Canada}
\affiliation[b]{Perimeter Institute for Theoretical Physics, Waterloo, Ontario, N2L 2Y5, Canada }
\emailAdd{cburgess@perimeterinstitute.ca} 
\emailAdd{haingej@mcmaster.ca}
\emailAdd{kaplaneg@mcmaster.ca}
\emailAdd{rummelm@mcmaster.ca}

\date{}

\abstract {Persistent puzzles to do with information loss for black holes have stimulated critical reassessment of the domain of validity of semiclassical EFT reasoning in curved spacetimes, particularly in the presence of horizons. We argue here that perturbative predictions about evolution for very long times near a horizon are subject to problems of secular growth -- {\it i.e.}~powers of small couplings come systematically together with growing functions of time. Such growth signals a breakdown of naive perturbative calculations of late-time behaviour, regardless of how small ambient curvatures might be. Similar issues of secular growth also arise in cosmology, and we build evidence for the case that such effects should be generic for gravitational fields. In particular, inferences using free fields coupled only to background metrics can be misleading at very late times due to the implicit assumption they make of perturbation theory when neglecting other interactions. Using the Rindler horizon as an example we show how this secular growth parallels similar phenomena for thermal systems, and how it can be resummed to allow late-time inferences to be drawn more robustly. Some comments are made about the appearance of an IR/UV interplay in this calculation, as well as on the possible relevance of our calculations to predictions near black-hole horizons. 
}

\begin{document}

\maketitle
\section{Introduction and summary}

This paper argues that, for quantum fields in the presence of horizons, perturbation theory is like good weather: if you wait long enough it eventually fails. This might be important to the extent that the properties of free fields in curved spacetime are used to draw inferences about late-time behaviour (as is often done for near-de Sitter cosmology and black hole information loss). 

Quantum field theory on spacetimes with horizons turns out to contain many surprises. The first of these was the discovery of Hawking radiation \cite{Hawking:1974sw, Wald:1975kc} and the peculiar relationship between gravity and thermodynamics \cite{Bekenstein:1973ur, Bardeen:1973gs, Bekenstein:1974ax, Hawking:1976de, Gibbons:1976ue}. This is a gift that even now continues giving, leading to many associated puzzles about information loss that cut to the foundations of quantum mechanics and general relativity \cite{Hawking:1976ra, Almheiri:2012rt} (for reviews with more extensive references, see \cite{
Banks:1994ph, Mathur:2009hf, Harlow:2014yka, Polchinski:2016hrw, Marolf:2017jkr}). It is perhaps noteworthy that many of these puzzles involve understanding the system at very late times, such as the Page time \cite{PageTime1, PageTime2} after which significant amounts of information are radiated.

What is striking is that these puzzles seem to rely simply on the existence of horizons and do not require large curvatures. As such they seem to be accessible entirely within the domain of validity of effective field theory (EFT) methods \cite{Weinberg:1978kz} (for reviews of these techniques applied to gravity, see \cite{ EFTgrav1, EFTgrav2, EFTgrav3}). As a consequence they appear to arise within controlled calculations that should accurately capture the quantum system's behaviour. Furthermore, couching them within an EFT framework ensures they do not rely on the details concerning whatever unknown physics unitarizes quantum gravity at very high energies. Indeed, the persistence of these puzzles has stimulated re-examination of the domain of validity of EFT techniques, and a reassessment of whether these methods might sometimes unexpectedly break down near horizons even when curvatures are small.

\subsection*{Thermality, open systems and late-time breakdown of perturbative methods}

When identifying the domain of validity of EFT methods near horizons a possibly important observation is that quantum systems near horizons resemble open thermal systems in two related ways. First, they are thermal inasmuch as the spectrum of emitted radiation takes a black-body form (up to grey-body factors). Second, they are open in the sense that the existence of the horizon ensures there is an unobserved sector ({\it e.g.}~the black-hole interior) that entangles and exchanges information with the observers of interest (those outside the black hole).

The observation that these are open systems can be germane to identifying the domain of validity of EFT methods. Wilsonian EFT reasoning is usually assumed to apply provided only there exists an appropriate hierarchy of energy scales, $E_1 \ll E_2$, with the EFT defined (by construction) by the action built only from low-energy degrees of freedom that reproduces the physics of some more UV-complete system order-by-order in the small ratio $E_1/E_2$. But for open thermal systems it can sometimes happen that having a small parameter is insufficient to justify perturbative methods, regardless of how small the relevant parameter turns out to be.

To see why, recall that open systems by definition are those for which a sector of the system (called here the `environment') is not measured in detail. In the special case of a thermal system this environment consists of all those parts of the heat bath that are ignored and replaced by a coarse-grained thermodynamic description. It is the tracing over these unmeasured degrees of freedom that allows the rest of the system to be described by a density matrix rather than a pure state. But if an environment like this is always present, it is generic that it can cause a breakdown in perturbative methods at late times. 

This late-time breakdown of perturbation theory has it roots in the observation that the time-evolution operator is $U(t) = \exp(- iHt)$, and no matter how $H$ gets split into perturbed and unperturbed parts -- {\it i.e.} $H = H_0 + H_{\rm int}$ -- it is always true that there is eventually a time late enough that it is a bad approximation to compute $U(t)$ in powers of $H_{\rm int} t$. Put another way: no matter how weak the coupling to the environment, if the environment never goes away its eternal presence can eventually allow perturbatively small effects to accumulate and become large.\footnote{A familiar example of this occurs with the passage of light through a transparent medium. Regardless of how weak the interactions are between photons and their environment (the medium), one eventually transitions to a regime where the light either reflects or refracts. Since neither of these options need be a small change to the direction of the incident light, a description of this perturbing in the light-medium interaction is destined eventually to break down.} This kind of late-time breakdown of perturbation theory --- often called secular growth, due to the appearance of increasing functions of time order-by-order in the perturbative expansion --- is less familiar in particle physics, for which the focus is usually on scattering and so interactions get turned off at late times as wave packets separate from one another.

Of course the existence of secular growth need not completely preclude making late time predictions. For instance, one believes in exponential decay laws, 
\be \label{exponential}
 n(t) = n_0 \, e^{-\Gamma t} \,, 
\ee
for times well after the mean lifetime ({\it i.e.} for $\Gamma t \gg 1$) even though the decay rate $\Gamma$ is usually computed in perturbation theory. The exponential decay is robust in this sense because it relies only on the evolution of the number of particles satisfying $\exd n/\exd t = - \Gamma n$ for all times, since the integration of this evolution ensures the validity of \pref{exponential}, regardless of how large $\Gamma t$ might be. Observations like this ultimately underly the ability to resum late-time behaviour for thermal systems, but even when resummations are possible the existence of secular evolution has implications (such as by giving physical predictions that are not analytic in the small couplings, sometimes making them not captureable by a simple Taylor series).

Thermal systems provide among the best-studied examples of open systems and there are indeed well-known examples where perturbation theory does break down for thermal systems. One pertinent to what follows is the need to resum thermal mass effects --- `hard thermal loops' --- for massless bosons in thermal environments \cite{Gross:1980br, Braaten:1989mz}. Another, related, example is the well-known breakdown of perturbative (mean-field) methods when calculating near a critical point \cite{CritBreakdown}. Both of these examples are also associated with infrared effects due to the presence of massless (or very light compared with the temperature) bosons. Mathematically, the fluctuations of these bosons tend to cause correlation functions to become singular at low frequencies and this can both give rise to infrared (IR) divergences and to the secular growth and divergence of these correlation functions at late times. These fluctuations are stronger at nonzero temperature because of the singular small-$k$ limit of the Bose-Einstein distribution function, 
\be
    f_\ssB(k) = \frac{1}{e^{k/T}-1} \simeq \frac{T}{k} \left[ 1 + \cO\left(\frac{k}{T} \right)\right] \,, 
\ee
when evaluated for dispersion relation $\omega(k) = k$.

\subsection*{Relevance for QFT in curved space}

The thermal nature of quantum systems near horizons combined with the breakdown of late-time perturbation theory that can occur for thermal systems makes it natural to ask whether or not perturbation theory also can break down at late times near horizons. If so, then the mere existence of a small expansion parameter (such as the low-energy derivative expansion underlying EFT methods) is insufficient to justify some calculations; in particular those relevant to very late times. This would represent an ill-appreciated boundary to the domain of validity of semiclassical methods for quantum fields in curved space.  

This is particularly important for applications to gravity, for which the very validity of semiclassical methods relies on EFT methods that reveal semiclassical perturbation theory to be in essence a derivative expansion. Semiclassical methods are predictive only to the extent that quantum loops are suppressed by successively higher powers of the low energies and curvatures, divided by the UV scales of the problem (such as, but not restricted to, the Planck mass, $M_p$).

If open systems are a guide then having a good hierarchy of scales need not in itself be sufficient to justify perturbation theory at late times. This is because regardless of how small this perturbative parameter is, there is a danger that powers of this parameter enter into observables systematically multiplied by growing powers of time. If this occurs (and it can occur for thermal systems), no matter how small the interaction, there is always a time after which it can no longer be treated perturbatively. 

Notice that, if present, this kind of late-time breakdown of perturbation theory could be important for arguments like those leading to inferences about very late times during inflation or about information loss near black holes. This is because these arguments tend to be based on calculations involving free fields moving on curved spacetime, for which the neglect of interactions amounts to the assumption that these interactions can be treated perturbatively. If QFT near horizons share the late-time breakdown of perturbation theory experienced by open systems then this basic assumption could be at risk. 

Even so, controlled conclusions at late times might yet be possible, although some sort of resummation might be necessary in order to do so. Indeed there is evidence that such resummations are possible for cosmological horizons, for which quantum effects in multi-field models are known to exhibit secular growth that undermines late-time perturbative calculations \cite{CosmoSecular1, CosmoSecular2, CosmoSecular3, CosmoSecular4, CosmoSecular5, CosmoSecular6, CosmoSecular7, CosmoSecular8, CosmoSecular9, CosmoSecular10, CosmoSecular11, CosmoSecular12, CosmoSecular13, CosmoSecular14, CosmoSecular15, CosmoSecular16, CosmoSecular17, CosmoSecular18, CosmoSecular19}. In this case the resummed late-time evolution \cite{TsamisWoodard, FinelliMarozzi} appears to be given by the stochastic formulation of inflationary theories \cite{Stochastic1, Stochastic2}, plus calculable corrections \cite{OpenEFT2}. As a bonus, the recognition that these are essentially open systems also leads to an understanding of why initially quantum fluctuations decohere while outside the horizon, re-entering as the primordial classical fluctuations used to describe the later universe \cite{OpenEFT1, Decoherence1, Decoherence2, Decoherence3, Decoherence4, Decoherence5, Decoherence6, Decoherence7, Decoherence8, Decoherence9, Decoherence10, Decoherence11, Decoherence12}.

\subsection*{This paper}

In this paper we start --- see however \cite{Akhmedov:2015xwa} --- the process of assessing secular behaviour near black hole horizons at late times, starting with uniformly accelerated (Rindler) observers in flat space. This example has the advantage that the connection between acceleration and thermality is very well-understood \cite{Unruh:1976db, Unruh:1983ac, Lee:1985rp, Takagi:1986kn, Troost:1977dw} and calculations are explicit. 

We argue here that Rindler observers working within the Minkowski vacuum indeed do see the instances of secular growth and perturbative breakdown that one would expect given their thermal nature. In particular by studying a massless scalar field $\phi$ subject to a $\lambda \phi^4$ interaction on flat space we show how Rindler observers run into difficulty naively perturbing in powers of $\lambda$ at late Rindler times. This happens in precisely the way expected based on the Rindler observer's thermal interpretation of the Minkowski vacuum. Interestingly, the same does {\it not} happen for a Minkowski observer because of an IR/UV interplay: they find differing results at late time because the two kinds of observers do not use the same renormalization for very short-distance fluctuations. 

Our story is organized as follows. First, \S\ref{sec:secular} reviews the appearance of power-law secular growth, starting in \S\ref{ssec:Thermal} at late Minkowski time for two-point correlation functions of thermal systems. \S\ref{ssec:Rindler} then repeats the calculation for correlation functions computed using the zero temperature Minkowski vacuum, both showing why the finite-temperature effects go away for a Minkowski observer, but instead persist (though at late Rindler time) for a Rindler observer. The Rindler observer also sees power-law growth with time, with a coefficient agreeing with the thermal result if the temperature is given by the Unruh result $T_a = a/2\pi$ \cite{Unruh:1976db}. The main difference between Minkowski and Rindler results arises because these make different choices for the two-point mass counterterm. The Rindler observer makes a different subtraction in order to avoid late-time secular evolution for the Rindler ground state. Once the Rindler observer subtracts to ensure no such growth for correlations in the Rindler vacuum the result in the Minkowski vacuum is completely fixed, leading to IR effects not present otherwise.  

This explanation of secular growth is followed, in \S\ref{sec:NHResum}, by a discussion of how the spurious secular growth can be resummed to obtain reliable late-time predictions. Although a general formalism for this often exists for open systems \cite{OpenEFT2}, its use is over-kill in this particular instance since the basic breakdown in late-time perturbation theory arises because of a nonzero mass shift. Resumming mass shifts is easy to do since this is what the Schwinger-Dyson equations tell us how to do, and in the present instance simply amounts to replacing the naive leading-order propagator with the one corresponding to the appropriate mass.

Once this resummation is done the resulting two-point function falls off to zero for large invariant separation, $s$. It does so more slowly, however, than does the leading-order propagator, falling like $s^{-3/2}$ rather than $s^{-2}$ at large times. It is this relative growth of the difference between the accurate propagator and its leading approximation to which the secular breakdown of perturbation theory points. 

Some further conclusions are summarized in \S\ref{sec:Conclusions}, including preliminary discussions of the potential relevance of this calculation to late-time black-hole physics.

\section{Secular growth and perturbation theory}
\label{sec:secular}

A scalar field in curved spacetime is the poster child for the interplay of quantum effects with gravitational fields, with most discussions being limited to free quantum fields propagating on a background metric. Implicit in such discussions is the justification for neglecting any other interactions amongst the quantum fields. To study this justification more explicitly we therefore explore the implications of self-interactions for a real quantum scalar field, $\phi(x)$, using this as our vehicle for tracking late-time issues in perturbation theory.

To this end consider the following action describing the self-interaction of a real scalar quantum field $\phi(x)$ in the presence of a background gravitational field, $g_{\mu\nu}$:
\be
 S = - \int \exd^4 x \, \sqrt{-g} \; \left[ \frac12 \, g^{\mu\nu} \partial_\mu \phi \, \partial_\nu \phi + V(\phi) \right] \,,
\ee
with scalar potential chosen to describe self-interactions of massless particles
\be
  V(\phi) = \frac{\lambda}4 \, \phi^4 \,.
\ee
We choose small $\lambda$ and our interest lies in the validity of perturbative calculations in powers of $\lambda$. 

In any real system many other interactions typically also exist, including interactions between $\phi$ and quantum fluctuations of the metric, and these can be treated using standard EFT methods \cite{EFTgrav1,EFTgrav2,EFTgrav3}. Although we focus for simplicity purely on the self-interaction $\lambda \phi^4$, we believe our treatment applies equally well to any other perturbatively small interactions in this type of EFT framework.

\subsection{Thermal calculation}
\label{ssec:Thermal}

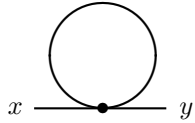
\begin{figure}
\centerline{
\begin{picture}(260,100)
\put(100,0){\begin{picture}(100,100)
    \thicklines
   \put(36,40){\circle{40}}
   \put(10,20){\line(1,0){50}}
   \put(36,20){\circle*{4}}
   \put(0,18){$x$}
   \put(65,18){$y$}
    \end{picture}}
\end{picture}
}
\caption{Feynman graph giving the leading one-loop `tadpole' correction to the scalar propagator. The result obtained by evaluating this graph must be summed with the tree-level two-point counter-term graph.\label{figure:Tadpole}}
\end{figure}

Let us first display how secular growth arises for this scalar-field system, and how this can undermine the validity of perturbative expansions in powers of $\lambda$. To do so consider preparing the system in a thermal state in Minkowski space, $g_{\mu\nu} = \eta_{\mu\nu}$, with density matrix
\be \label{ThermalState}
 \rho = \frac{1}{Z} \, \exp \Bigl( - H/T \Bigr) \,.
\ee
Here $Z$ is chosen (as usual) to ensure Tr $\rho = 1$ and $H$ is the conserved Hamiltonian generator of time-translations in Minkowski space,
\be
 H = \int \exd^3 x \, \left[ \frac12 \, (\partial_t \phi)^2 + \frac12 \, (\nabla \phi)^2 + V(\phi) \right] \,,
\ee
with the integration taken over a spatial slice at fixed $t$. 

Since our interest is in secular evolution and perturbative corrections we compute order-$\lambda$ corrections to the correlation function $\cG(x; y) = \langle T \phi(x) \phi(y) \rangle$, where $T$ denotes time-ordering and the average is taken using the thermal state \pref{ThermalState}: $\langle \cO \rangle = \hbox{Tr }(\rho\, \cO)$. Since time-dependence is kept (as opposed, say, to thermodynamic properties) the correlation function is computed using the Keldysh formalism \cite{Schwinger:1960qe, Keldysh:1964ud}. 

\subsubsection{Secular growth}

The leading correction to $\cG(x;y)$ in powers of $\lambda$ arises at one loop, and is obtained by evaluating the tadpole graph of Fig.~\ref{figure:Tadpole}. The loop part of this graph diverges in the UV and is renormalized as at zero temperature (such that the renormalized zero-temperature mass remains zero), leaving the usual finite-temperature residual mass shift
\be \label{dmsqT}
  \delta m_\ssT^2 = \frac{ \lambda T^2}{4} \,.
\ee
Using this self-energy in the remainder of the graph, and evaluating the result at zero spatial separation leads to the following form in the limit of large temporal separation:
\be \label{Tsect}
  \cG_{\rm tad}(x^0, \bfx; y^0 = x^0+t, \bfy = \bfx) = \frac{\delta m_\ssT^2 \,T t}{8\pi} + \cdots = \frac{\lambda T^3 t}{32 \pi} + \cdots\,,
\ee
where the ellipses denote terms growing more slowly with $t$ at large temporal separation. Details of this calculation are provided in Appendix \ref{App:ThermCalc}. 

Eq.~\pref{Tsect} provides an example of secular growth: regardless of how small $\lambda$ may be, the theory tells us that perturbation theory built around it eventually breaks down for late enough times. Later sections give more details about why this occurs, but for now it suffices to mention that this type of late-time breakdown of perturbative methods is very common in most areas of physics (though not so much in particle physics).

\begin{figure}
\centerline{
\begin{picture}(260,100)
\put(100,0){\begin{picture}(100,100)
    \thicklines
   \put(36,40){\circle{40}}
   \put(36,80){\circle{40}}
   \put(10,20){\line(1,0){50}}
   \put(36,20){\circle*{4}}
   \put(36,60){\circle*{4}}
   \put(0,18){$x$}
   \put(65,18){$y$}
    \end{picture}}
\end{picture}
}
\caption{Feynman graph giving a subleading two-loop `cactus' correction to the scalar propagator. Although this is not the only two-loop contribution, it is noteworthy because of the power-law (as opposed to logarithmic) IR divergences it acquires due to the singularity of the Bose-Einstein distribution at small momenta. \label{figure:Cactus}}
\end{figure}
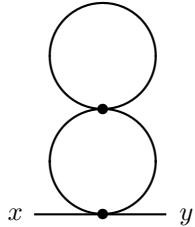

To summarize, late-time perturbative breakdown typically arises whenever there is a persistent environment with which a system interacts. In such a case late times are a problem because no matter how weak the interaction with the environment might be, eventually the secular accumulation of arbitrarily small effects can give big results. Eq.~\pref{Tsect} provides an illustration of this effect, where it is the relevant persistent environment is in this case the thermal bath itself.

\subsubsection{IR divergences}

Secular growth at late times is related to (but not identical with) infrared divergences when performing loops involving intermediate states with arbitrarily small frequency. This connection arises because divergence of a correlation function at late times implies singularity of its Fourier transform at small frequencies, and such singularities can also cause loop integrals to diverge because of contributions near $\omega = 0$. 

Although the graph of Fig.~\ref{figure:Tadpole} is not infrared divergent, massless bosons can also experience enhanced IR divergences at finite temperature. A simple example of this can be seen once two-loop corrections to the scalar propagator are examined, such as the `cactus' graph of Fig.~\ref{figure:Cactus}. In this case the upper loop again contributes the temperature-dependent mass shift given in \pref{dmsqT}, while the lower loop diverges like a power of an IR cutoff, $\omega_{\rm min}$, contributing a factor
\be
  \hbox{lower loop} \propto \frac{\lambda T}{\omega_{\rm min}} \,.
\ee
At zero temperature this lower loop would diverge logarithmically, $\propto \int \exd^4k/(k^2)^2$, but this becomes a power divergence at finite temperature because of the singularity of the Bose-Einstein distribution,\footnote{Alternatively, if evaluated within the Euclidean formulation the extra divergence arises because the integration is only over spatial components, $\exd^3 k$, because the frequency integral is replaced by a Matsubara sum.} $n_\ssB(k) = (e^{k/T}-1)^{-1} \simeq T/k$ for $k \ll T$.

These kinds of IR divergences also signal a breakdown of perturbative methods, in this case due to the unsuppressed contribution of frequencies $\omega \le \lambda T$ in loops, where the naive suppression by powers of small coupling can instead systematically become powers like $(\lambda T/\omega)^L$. The need to resum\footnote{Although controlled resummation is possible for massless scalars it need not always be true that such IR issues can be solved by resumming specific classes of higher-order graphs. An example where things are more serious arises when the zero-temperature squared mass is adjusted to be slightly negative, $m_0^2 < 0$, so that it is the {\it total} mass, $m^2 = m_0^2 + \delta m_\ssT^2$, that vanishes. This corresponds to arranging the theory to sit at a critical point, for which it is well-known that mean-field (perturbative) calculations are simply not a good approximation.} the IR part of such graphs leads to well-known hard-thermal-loop effects whose dependence on fractional powers of $\lambda$ also reveals a breakdown in validity for naive expansions in integer powers of $\lambda$. 

We notice in passing that the zero-temperature limit of the graph of Fig.~\ref{figure:Cactus} does {\it not} contribute an IR divergence. This is at first a surprise because, as mentioned above, the bottom loop appears to contribute proportional to $\propto \int \exd^4k/(k^2)^2$ and so should diverge logarithmically in the IR. The total graph nonetheless does not diverge because the upper loop in Fig.~\ref{figure:Cactus} (or the only loop in Fig.~\ref{figure:Tadpole}) is renormalized to ensure $m^2 = 0$ at zero temperature, ensuring that the naive zero-temperature graph simply cancels with the mass counter-term. The same cancellation does not occur at finite temperature because the counter-terms only cancel the zero-temperature part of the top tadpole contribution. 

This relevance of a UV renormalization to the IR behaviour of the result is an example of how UV/IR mixing plays a role in our subsequent calculations.

\subsection{Rindler calculation}
\label{ssec:Rindler}

\begin{figure}[t]
\begin{center}
\includegraphics[width=30mm,height=60mm]{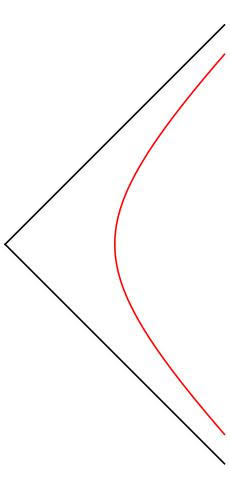} 
\caption{Cartoon of a Rindler wedge, showing both the (red) hyperbolic world-line of an accelerating observer and the (black) horizons to which it asymptotes.} \label{fig:Wedge} 
\end{center}
\end{figure}

The previous section illustrates several ways for which IR issues can make perturbative expansion for thermal bosons be more subtle than one might think based purely on zero-temperature calculations. The existence of these types of IR issues then raises a potential puzzle: even zero-temperature scalar field theory on flat space can be thought of as a thermal state from the point of view of a Rindler observer who is uniformly accelerated relative to inertial observers. 

If a Rindler observer sees the usual Minkowski vacuum as a thermal state, the ordinary zero-temperature Minkowski propagator should display thermal perturbative secular growth as seen by a Rindler observer. How does this come about? And if it does come about, why isn't secular growth and perturbative breakdown also a problem for Minkowski observers?

These questions are clarified in this section, where we show how the ordinary vacuum indeed does include secular growth, at least when renormalized as appropriate for a Rindler observer and when evaluated at late {\it Rindler} times. As we show below the potential breakdown of perturbative methods associated with thermal secular growth turns out to coincide with a breakdown of perturbation theory that is also expected at zero temperature.

To proceed we reproduce the above calculations at zero temperature, and so evaluate the correlation function $G(x;y) = \langle 0 | T \phi(x) \phi(y) | 0 \rangle$ for the standard Minkowski vacuum, $|0\rangle$. The corrections to this correlation function arise by evaluating the same graphs as before, but using standard zero-temperature Feynman rules. 

For later purposes it is useful to have explicit expressions for the lowest-order propagators when the scalar mass is nonzero, and this is given by
\be \label{G0prop}
  G_0(x;y) = -i \int \frac{\exd^4p }{(2\pi)^4 } \; \frac{e^{ip\cdot(x-y)}}{p^2 + m^2 - i \epsilon} = \frac{1}{4\pi^2} \;\frac{m}{\sqrt{(x-y)^2 + i\epsilon}} \; K_1\Bigl[ m \sqrt{(x-y)^2 + i \epsilon} \Bigr] \,,
\ee
where in our $(-+++)$ conventions $(x-y)^2 = \eta_{\mu\nu} (x-y)^\mu (x-y)^\nu$ is negative for time-like separations and positive for space-like separations. Here $K_\nu(z)$ is a modified Bessel function, for which we record the following useful asymptotic forms \cite{Abr, Watson}
\be
  K_1(z) \simeq \frac{1}{z} + \left[ -1 + 2 \gamma + 2 \ln \left( \frac{z}{2} \right) \right] \frac{z}4 +\cO(z^3) \qquad \hbox{for $|z| \to 0$}  \,, 
\ee
(with $\gamma$ the Euler-Mascheroni constant) and
\be
 K_1(z) \simeq \sqrt{ \frac{\pi}{2z}} \; e^{-z} \left[1 + \frac{3}{8z} + \cdots \right] \qquad \hbox{for $|z| \to \infty$ if $|\mathrm{arg}(z)| < \tfrac{3\pi}{2}$}\,.
\ee
In particular, the massless limit of \pref{G0prop} gives
\be
 G_0(x;y) =  -i \int \frac{\exd^4p }{(2\pi)^4 } \; \frac{e^{ip\cdot(x-y)}}{p^2 - i \epsilon} = \frac{1}{4\pi^2} \; \frac{1}{(x-y)^2 + i\epsilon} \qquad (\hbox{$m=0$}) \,.
\ee

\subsubsection{Secular growth}

With these expressions in hand we may now evaluate the tadpole graph of Fig.~\ref{figure:Tadpole}. Evaluating the self-energy loop factor --- {\it i.e.}~the part of the graph without the external lines --- gives a momentum-independent result which we denote as\footnote{In our conventions the momentum-space propagator is $-i/(p^2+m^2 - \Sigma)$.} $i\Sigma(0)$, and whose value is given by
\be
  \Sigma(0) = -\delta m^2_{\rm ct} + 3i  \lambda \int \frac{\exd^4 k}{(2\pi)^4} \; \frac{1}{k^2 -i \epsilon} \,,
\ee
where $\delta m^2_{\rm ct}$ is the mass counter-term that subtracts the UV-divergent part of the integral. We return to evaluate this expression in more detail later. 

Temporarily re-installing a mass for $\phi$, the remainder of the graph evaluates to give
\ba \label{Gtad1loop}
  G_{\rm tad}(x;y) &=& -i \Sigma(0) \int \frac{\exd^4p}{(2\pi)^4} \; \frac{ e^{i p \cdot (x-y)}}{(p^2 + m^2 - i \epsilon)^2} = \Sigma(0) \left(- \frac{\partial}{\partial m^2} \right) G(x;y) \nn\\
  &=& -\frac{\delta m^2}{8\pi^2}  \; K_0\Bigl[ m \sqrt{(x-y)^2 + i \epsilon} \Bigr] \,,
\ea
where $\delta m^2 = -\Sigma(0)$ is the UV finite mass shift after renormalization. In the massless limit this evaluates to 
\be \label{MinkTadxy}
  G_{\rm tad}(x;y) \to  \frac{\delta m^2}{8\pi^2} \; \ln \Bigl[\mu \sqrt{(x-y)^2 + i \epsilon} \Bigr] \,.
\ee
up to a spacetime-independent IR-divergent constant. The precise value of the scale $\mu$ depends on how this IR divergence is regulated, though we do not pursue this because its value plays no role when tracking the dependence of the result on $x-y$. Notice that for massless fields \pref{MinkTadxy} holds for all $x$ and $y$ and not just asymptotically in the limit $(x-y)^2 \to 0$. 

The above agrees at leading nontrivial order with the standard operator-product expansion \cite{OPE1, OPE2}
\be
 \phi(x) \, \phi(0) \sim C(x) + D(x) \, \phi^2(0) + \cdots \qquad \hbox{(as $x^2 \to 0$)} \,,
\ee
where the RG-improved one-loop expressions for the coefficients are
\be \label{RGimprove1}
  C(x) = \frac{1}{4\pi^2 x^2} -\left( \frac{m^2}{3\lambda} \right) D(x)[1-D(x)] \quad \hbox{and} \quad
  D(x) = (1+L)^{1/3} \,,
\ee
with
\be \label{RGimprove2}
 L := \frac{9 \lambda}{16 \pi^2} \, \ln \Bigl( \mu^2 x^2 \Bigr) \,.
\ee
Here $\mu$ is the renormalization scale for the renormalized coupling $\lambda(\mu)$. 

Imagine now choosing both $x^\mu$ and $y^\mu$ to lie within the same Rindler wedge, for which we adopt coordinates $(\tau, \xi)$ defined relative to $(t,x) = (x^0, x^1)$ by
\be
  x = \xi \cosh(a \tau) \qquad \hbox{and} \qquad t = \xi \sinh(a \tau) \,,
\ee
in terms of which the Minkowski metric becomes
\be
  \exd s^2 = - \eta_{\mu\nu} \exd x^\mu \exd x^\nu = - (a\xi)^2 \exd \tau^2 + \exd \xi^2 + \exd y^2 + \exd z^2 \,.
\ee

Next evaluate $G(x,y)$ with $x^2 = y^2$ and $x^3 = y^3$ and choose a particular Rindler observer -- {\it i.e.}~fixed $\xi$ --- on whose accelerating world-line both $x^\mu$ and $y^\mu$ lie. From the point of view of this Rindler observer these two points are separated purely by a shift in Rindler time, $\tau$, and this is the proper time along this trajectory if $\tau$ is rescaled so that $a = 1/\xi$. In this case $a$ also represents the proper acceleration of this trajectory.

The invariant separation between these points is then related to their Rindler time difference by
\be
 - (x-y)^2 = \frac{4}{a^2} \; \sinh^2 \left( \frac{a \tau}2 \right) \simeq \frac{1}{a^2}\, e^{a \tau} \Bigl[ 1 + \cO(e^{-2a\tau}) \Bigr] \,,
\ee
where the final approximate equality gives the asymptotic form when $x$ and $y$ are chosen so that $a \tau \gg 1$. For such an observer, when $a \tau \gg 1$ eq.~\pref{MinkTadxy} implies an asymptotic secular growth of $G_{\rm tad}(\tau)$ with $\tau$ of the form
\be \label{GRind1}
 G_{\rm tad}(\tau) =  \frac{\delta m^2}{8\pi^2} \; \ln \Bigl[\mu \sqrt{(x-y)^2 + i \epsilon} \Bigr] \simeq  \frac{\delta m^2 \, a \tau}{16\pi^2} + \hbox{subdominant} \,.
\ee

What value should be chosen for $\delta m^2$ in this last expression? As discussed earlier, the tadpole loop diverges in the UV and a Minkowski observer would renormalize this divergence by cancelling it against a mass counter-term, thereby effectively choosing $\delta m^2_\ssM =  - \Sigma_\ssM(0) = 0$. A Rindler observer would also have to choose a counter-term, and the natural choice for this observer would be to demand the sum of counterterm and tadpole graph vanish if evaluated in the Rindler ground state.\footnote{The Rindler state is the ground state of the Rindler hamiltonian, defined as the Poincar\'e boost generator that generates translations in Rindler time.} Said differently, secular growth at late Rindler time cannot be avoided for {\it both} the Minkowski and Rindler vacua, and once it is excluded from the Rindler vacuum it necessarily appears for the Minkowski state, and does so in precisely the way expected for thermal secular growth.

Once this is done the counter-term no longer completely cancels the tadpole graph when it is evaluated in the Minkowski vacuum, and so differs from the Minkowski choice by a finite amount. A standard calculation using the Rindler vacuum gives a finite residual result \cite{Davies:1974th, Boulware:1974dm, Troost:1978yk, Dowker:1978aza, Takagi:1986kn, Linet:1995mq}
\be \label{amass}
 \delta m^2_a  = \frac{ \lambda a^2}{16\pi^2} \,.
\ee
Using this in \pref{GRind1} then gives
\be
 G_{\rm tad}(\tau) =  \frac{\lambda a^3 \tau}{(16\pi^2)^2} + \hbox{subdominant} \,,
\ee
which agrees with the thermal result, \pref{Tsect}, provided we identify temperature with acceleration in the usual way:
\be
  T_a = \frac{a}{2\pi} \,.
\ee

\subsubsection{IR divergences}

A similar discussion applies when describing the power-law IR divergences encountered in the cactus graph of Fig.~\ref{figure:Cactus}. The divergence of this graph in the IR is one of the signals of perturbative breakdown for thermal systems \cite{Cactus1, Cactus2}. Once the top loop in this graph is recognized to be nonzero ({\it i.e.}~$\delta m_a^2 \ne 0$ because of the Rindler renormalization choice), the bottom loop diverges in the infrared. If evaluated in position space this loop involves a factor
\be \label{CactusPosn}
\hbox{lower loop} \propto \int  \exd^4 v \; [ G(u;v) ]^2 = \frac{1}{(2\pi)^4} \int \frac{ \exd^4 v }{[(u-v)^2 + i \epsilon]^2} \,,
\ee
where $v$ denotes the interaction point in Fig.~\ref{figure:Cactus} where the two loops touch.

The IR divergence in this language corresponds to a failure of \pref{CactusPosn} to converge in the limit $v \to \infty$. This divergence can be regularized by imposing an upper limit on the absolute value of the invariant separation $L =| (u-v)^2|_{\rm max}$, and the divergence is logarithmic in this invariant cutoff. The IR divergence becomes a power-law divergence once $L$ is re-expressed in terms of a maximum Rindler time, $\tau_{\rm max}$, since $L$ is exponential in $\tau_{\rm max}$. 

\subsubsection{Other graphs}

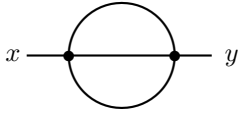
\begin{figure}
\centerline{
\begin{picture}(260,100)
\put(100,0){\begin{picture}(100,100)
    \thicklines
   \put(36,40){\circle{40}}
   \put(0,40){\line(1,0){70}}
   \put(16,40){\circle*{4}}
   \put(56,40){\circle*{4}}
   \put(-8,38){$x$}
   \put(75,38){$y$}
    \end{picture}}
\end{picture}
}
\caption{Feynman graph giving a two-loop correction to the scalar propagator that is not proportional to the one-loop tadpole contribution of Fig.~\ref{figure:Tadpole}. \label{figure:SettingSun}}
\end{figure}

So far all of the problematic graphs involve the basic tadpole, and as a consequence are proportional to $\delta m^2$. This makes them vanish or not depending on the counter-term choices made when renormalizing the tadpole's UV divergence. Do all secular effects disappear if $\delta m^2$ vanishes?

To address this requires examining other graphs, such as the `setting-sun' diagram of Fig.~\ref{figure:SettingSun}. The self-energy contribution to this graph is given by the contributions excluding the external lines, and so evaluates to \cite{Sunset1, Sunset2}
\ba
  i\Sigma(k) &=& -i\lambda^2 \int \frac{\exd^4 p}{(2\pi)^4} \, \frac{\exd^4 q}{(2\pi)^4} \; \frac{1}{(p^2 - i \epsilon)(q^2 - i \epsilon)[(p+q+k)^2 - i \epsilon]} \nn\\
  &\simeq& A + B k^2 + C k^2 \ln\left( \frac{k^2}{\mu^2} \right) + \cdots\,,
\ea
where the second line takes the small-$k$ limit, where $A$ and $B$ are divergent constants and $C$ is calculable and finite.

Once external lines are attached this leads to a dependence on invariant separation $(x-y)^2$ of the form
\be
 G_{\rm sun}(x;y) = -i \int \frac{\exd^4 k}{(2\pi)^4} \; \frac{\Sigma(k)}{(k^2 - i \epsilon)^2} \; e^{ik\cdot (x-y)}\,.
\ee
On dimensional grounds this shows that it is the $k$-independent terms in $\Sigma$ that contribute to $G_{\rm sun}$ logarithmically in $(x-y)^2$, while those terms proportional to $k^2$ instead contribute proportional to $(x-y)^{-2}$. This suggests that it really is $\Sigma(0)$ that dominates for late Rindler times.

\section{Near-horizon resummation}
\label{sec:NHResum}

The previous section shows that, with the Rindler renormalization choice, the Minkowski vacuum shares the secular-growth phenomena associated with thermal states. In particular, perturbative corrections to Minkowski correlation functions involving massless bosons grow with Rindler time, indicating a breakdown of perturbation theory at large Rindler time. 

If perturbation theory breaks down in this regime how can the large-$\tau$ behaviour be computed? In this section we argue that it can be computed in the same way as is done for finite-temperature corrections: by resumming self-energy insertions to the propagator. 

The basic problem at late Rindler times arises because the Rindler observer chooses mass counter-terms that do not completely cancel the self-energy $\Sigma(k=0)$. As a result perturbation theory is organized in a way that obscures the correct dispersion relations, since these are not the poles of the lowest-order propagators. But the position of these poles is important when studying the limit where both $|\bfx - \bfy|$ and $x^0 - y^0$ are large and proportional to one another (as is the case for large Rindler time), because in this regime the propagator
\be
 G_0(x;y) = -i \int \frac{\exd^4p}{(2\pi)^4} \; \frac{e^{ip\cdot(x-y)}}{p^2 - i \epsilon} 
 = \int \frac{\exd^3p}{(2\pi)^3} \; \frac{1}{2|\bfp|} e^{-i|\bfp||x^0 - y^0| + i\bfp \cdot (\bfx - \bfy)}\,.
\ee
is dominated by the contributions of on-shell particles. But when particles are close to on-shell it is always a bad approximation to perturb in the self-energy, $\Sigma$, because it is the series
\be
  \frac{1}{p^2 - \Sigma} = \frac{1}{p^2} \sum_{n=0}^\infty \left( \frac{\Sigma}{p^2} \right)^n \,,
\ee
that is breaking down.  No matter how small $\Sigma$ is made (by choosing $\lambda$ small) there is always a regime near $p^2 = 0$ for which a series expansion in powers of $\lambda$ must fail. 

But this diagnosis also tells how to resum the terms that are not small in the perturbative expansion: one must use the full propagator for which all mass shifts are properly incorporated into the position of the propagator's pole. Happily for massless particles this can be read off from the form of the lowest-order massive propagator given in eq.~\pref{G0prop}, evaluated with $m^2 =\delta m^2_a$. 

The resummed result then is
\ba \label{G0prop2}
  G_{\rm rs}(x;y) &=& \frac{1}{4\pi^2} \;\sqrt{ \frac{\delta m^2_a}{(x-y)^2 + i\epsilon}} \; K_1\Bigl[  \sqrt{\delta m^2_a(x-y)^2 + i \epsilon} \Bigr] \nn\\
  &=& \frac{a}{16\pi^3} \; \sqrt{\frac{\lambda}{(x-y)^2 + i\epsilon}} \; K_1\left[ \frac{a}{4\pi}\, \sqrt{\lambda (x-y)^2 + i \epsilon} \right]  \,,
\ea
where the last equality evaluates $\delta m_a^2$ using eq.~\pref{amass}. Indeed for $|\delta m^2_a (x-y)^2| \ll 1$ this becomes
\be
 G_{\rm rs}(x;y) \simeq \frac{1}{4\pi^2} \left\{ \frac{1}{(x-y)^2 + i\epsilon} + \frac{\delta m^2_a}{4} \left[ -1 +2 \gamma + 2\ln \left( \frac{\sqrt{\delta m^2_a (x-y)^2 + i \epsilon}}2 \right) \right] + \cO[(\delta m^2_a)^2] \right\}\,.
\ee
The $\cO[\delta m^2_a]$ term in this expression precisely captures the dependence on $(x-y)^2$ found in \pref{MinkTadxy} by evaluating the tadpole graph. In this language the secular growth found when expanding in powers of $\delta m^2_a \propto \lambda$ simply reflects that the above expansion breaks down when $|\delta m^2_a (x-y)^2|$ is not small. 

But \pref{G0prop} also gives the result in the late-time limit when $|\delta m^2_a (x-y)^2|$ is not small. In this limit by using the large-argument limit of the Bessel function one instead finds 
\ba
  G_{\rm rs}(x;y) &\simeq& \frac{1}{\sqrt{32\pi^3}} \; \frac{(\delta m^2_a)^{1/4}}{[(x-y)^2 + i \epsilon]^{3/4}} \, \exp \left[ - \sqrt{\delta m^2_a (x-y)^2 + i\epsilon} \right] \nn\\
  &=& \frac{1}{8\pi^2} \sqrt{ \frac{a}{2}} \; \frac{\lambda^{1/4}}{[(x-y)^2 + i \epsilon]^{3/4}} \, \exp\left[- \frac{a}{4\pi}\, \sqrt{\lambda (x-y)^2 + i \epsilon} \right]  \,.
\ea
Notice this is {\it not} simply the same as using the operator-product expansion using RG-improvement, eqs.~\pref{RGimprove1} and \pref{RGimprove2}, to resum all orders in $\lambda \ln [m^2_a (x-y)^2]$. For time-like intervals, for which $(x-y)^2 = - (\Delta t)^2 < 0$, the exponential part of this last expression does not represent a suppression, leaving a result whose modulus falls like $(\Delta t)^{-3/2}$. For spacelike separations the suppression is instead exponential over a length scale set by $\sqrt{\delta m^2_a} = \sqrt\lambda \; a/4\pi$.

Notice that this predicts a falloff of the resummed result with temporal separation that is slower than for the lowest-order massless propagator, which instead satisfies $G_0 \propto (\Delta t)^{-2}$ for purely time-like separations. For this reason $G_{\rm rs}$ eventually is no longer a small change to $G_0$, and it is this breakdown of perturbation theory for which the secular growth was the signal. Notice also that the normalization of the late-time limit of $G_{\rm rs}$ is proportional to $(\delta m^2)^{1/4} \propto \lambda^{1/4}$, again reflecting a dependence on $\lambda$ that cannot be captured by a Taylor series about $\lambda = 0$.

\section{Discussion}
\label{sec:Conclusions}

This paper builds evidence for the picture advocated in \cite{OpenEFT1, OpenEFT2} that quantum fields in spacetimes with horizons are better described as open quantum EFTs than as traditional Wilsonian ones. Besides providing insights about how early quantum fluctuations can decohere \cite{OpenEFT1} in cosmology, this picture also suggests an unexpected breakdown of perturbative methods when computing late-time behaviour. 

Of course, nobody is surprised to hear that horizons lend themselves to treatment in terms of open systems. After all, it is the very essence of a horizon to cloak the physics in one region from the view of another, and this long ago suggested a formulation in terms of reduced density matrices \cite{Wald:1975kc, Hawking:1976ra}. What is less commonly appreciated is the generic threat the use of such techniques potentially raises for the use of semiclassical methods when discussing late-time behaviour. 

The problem with perturbative methods arises because for quantum fluctuations the gravitational field acts like an environment, and when this environment is static (or nearly so) its eternal presence provides unlimited time during which arbitrarily small perturbative effects can accumulate to cause large outcomes. This is a familiar phenomenon for particles interacting with a medium, and basically arises because no matter how small the interaction, $H_{\rm int}$, with the medium might be, there is always a time beyond which $\exp[{-iH_{\rm int} t}]$ differs significantly from $1- i H_{\rm int} t$. 

But this kind of breakdown of perturbative methods is particularly fatal for gravitational applications because semiclassical methods provide our only access to reliable predictions.\footnote{Notice that the issue here is {\it not} about the ability to solve Einstein's field equations only approximately, it is about the entire justification for thinking in terms of small quantum corrections to any classical solution in the first place.} Normally we think of semiclassical methods as relying on expanding in powers of derivatives of fields relative to the scale of gravitational UV completion (as outlined, for instance, in \cite{EFTgrav1, EFTgrav2, EFTgrav3}) and so feel justified in using them provided that spacetime curvatures remain small. But late-time perturbative breakdown sets a completely independent boundary to the domain of validity of semiclassical methods because it occurs regardless of how small the perturbation parameter is.

The evidence provided in this paper comes from the simplest situation having a horizon: an interacting scalar field probed by accelerated observers within flat spacetime. Such observers are known to regard the Minkowski vacuum as a thermal state, and based on this we search for evidence that the Minkowski vacuum displays the same kind of late-time secular effects that are found for thermal states. (Such effects arise when very light bosons interact with the heat bath, and in the simplest situations secularly growing perturbations arise from the same graphs that give such a boson a thermal shift to their mass.)

We find the same kinds of late-time secular growth arise for perturbative calculations of the two-point propagator in the (zero-temperature) Minkowski vacuum, provided both external points are restricted to lie along an accelerated trajectory within a single Rindler wedge, and provided that the mass counter-term is chosen so that late-time secular growth does not occur for the Rindler vacuum. The required counter-term expresses that the Rindler observer sees a finite acceleration-dependent mass shift (the direct analog of the temperature-dependent mass shift experienced by a particle interacting with a thermal bath). 

Perturbative breakdown occurs for the Minkowski state at late Rindler times, rather than late Minkowski time. It consequently occurs out towards null infinity, along the Rindler horizon. 

Because the secular growth has its roots in a nonzero mass shift, it is easy to resum: one merely organizes perturbation theory using the full shifted mass in the unperturbed part of the scalar field action. Once this is done the resummed propagator falls rather than growing at large times, but does so more slowly as a function of invariant separation than does the naive unperturbed propagator. From this point of view the secular growth seen in perturbation theory expresses the fact that the difference between the resummed and the naive propagators grows with time. 

\subsection{Relevance for black holes}

\begin{figure}[t]
\begin{center}
\includegraphics[width=90mm,height=60mm]{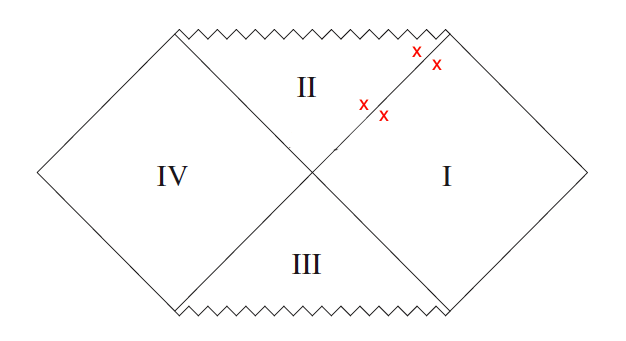} 
\caption{Penrose diagram for a Schwarzschild black hole, with crosses indicating the near-horizon points at which a correlation function $\langle \phi(x_1) \phi(x_2) \phi(x_3) \phi(x_4) \rangle$ might be evaluated to compute correlations relevant to information-loss arguments.} \label{fig:Penrose} 
\end{center}
\end{figure}

All this is suggestive that similar phenomena might also hold for black-hole spacetimes. Black holes are also believed to produce quantum fields in thermal states ---  with temperature set in terms of the Schwarzschild radius by  $T_\ssH \sim 1/r_s$. One might therefore expect the same kinds of secular growth and perturbative breakdown to occur for them as generically does for thermal states. These effects should be largest for bosons with masses much lighter than the Hawking temperature (such as for photons). 

For Schwarzschild black holes this would mean, for instance, that perturbations about a free quantum field prepared in the Hartle-Hawking state should show secular growth that ultimately causes perturbation theory to break down as one approaches future infinity along the black-hole event horizon. Even if these interactions are gravitationally weak, such as if $H_{\rm int} \sim G_\ssN T_\ssH^3 \sim T_\ssH^3/M_p^2 \sim \ell_p^2/r_s^3$ (where $G_\ssN \sim \ell_p^2 \sim 1/M_p^2$ is Newton's gravitational constant), then perturbative inferences would be expected to fail when 
\be
 t \gsim \frac{r_s^3}{\ell_p^2} \sim G_\ssN^2 M^3 \,,
\ee
which is of order the Page time \cite{PageTime1, PageTime2} associated with significant information loss. The time beyond which perturbative methods fail would be even earlier than this for interactions with dimensionless couplings larger than $T_\ssH^2/M_p^2$. Notice that this breakdown of perturbation theory can happen when the black hole is still very massive and so has nothing to do with the eventual appearance of large curvatures once $M \sim M_p$.

If the Rindler calculation is a reliable guide, the secular growth of perturbative interactions can be resummed to provide a reliable weak-coupling picture of what happens on such long time-scales. If singing from the Rindler scoresheet this resummation would correspond to shifting the mass by an amount $\delta m^2 \propto T_\ssH^2$, leading to a corresponding mass-dependent change in the asymptotic fall-off at late times.

Such a change in asymptotic large-time behaviour near the future horizon could well lead to large changes in inferences based purely on the quantum mechanics of free fields on a Schwarzschild background. For instance, consider a four-point correlation function $\langle \phi(x_1) \phi(x_2) \phi(x_3) \phi(x_4) \rangle$, evaluated with $x_i^\mu$ situated at the four points marked by crosses in the Penrose diagram of Fig.~\ref{fig:Penrose}. Calculating such a correlation function might be relevant to computing mutual entanglement between Hawking pairs at an early time and Hawking pairs at much later times. Any perturbative calculation of such a correlation function inevitably involves propagators evaluated with late-time separations in the danger zone near the future horizon, and so could well be poorly approximated by the free result. 

But what is the evidence that this kind of perturbation-violating secular growth actually occurs for black-hole spacetimes? Unfortunately, unlike for cosmology, very few explicit calculations exist, although suggestively there is one that does claim to find secular growth  \cite{Akhmedov:2015xwa}. 

An indication of how well intuition built on the Rindler example should apply to black holes can be found by performing the Rindler calculation in position space. In this case the tadpole expression evaluates to 
\ba \label{TadpolePosn}
G_{\rm tad}(x,y) &\propto& \delta m^2 \int  \exd^4 u \; G(x;u)\, G(u;y) = \frac{\delta m^2}{(2\pi)^4} \int \frac{ \exd^4 v }{[(v-x+y)^2 + i \epsilon][v^2+i\epsilon]}\nn\\
  &=& \frac{\delta m^2}{8\pi^2} \; \ln \Bigl[\mu \sqrt{(x-y)^2 + i \epsilon} \Bigr] + \hbox{(const)} \,.
\ea
This result --- {\it c.f.}~eq.~(\ref{MinkTadxy}) --- exhibits an $(x-y)$-independent IR divergence at large $v$, and converges for small $v$ so long as $(x-y)^2 \ne 0$. In this language it is the emergence of a divergence as $(x-y)^2 \to 0$ that is responsible for the logarithmic dependence on $(x-y)^2$ seen in the second line.

For arguments of secular growth it is the late-time limit $(x-y)^2 \to \infty$ that is of interest, and explicit calculation of the above integral shows this also to be logarithmic. What is important is this: the coefficient of the large-distance logarithm of interest for secular growth is identical to the coefficient of the short-distance logarithm in the limit $(x-y)^2 \to 0$.  This is important because this coefficient can therefore be determined purely from the integration region near $v^2 = 0$ and $(v-x+y)^2 = 0$; that is, by integration restricted to the near-horizon limit. 

To the extent that this is also true for black-hole spacetimes it would imply that it is only the near-horizon coincident limit of the propagator that matters for the late-time secular behaviour of perturbation theory. For general curved spaces this limit always has the Hadamard form $1/[4\pi^2 \sigma(x,y)]$ where $\sigma(x,y)$ is the invariant squared separation between $x$ and $y$. This suggests that the result could well agree with the result found using this form in the Rindler calculation. That this limiting form may suffice to understand the thermal features (and so late-time limit) of black holes also gets support from the observation that the coincident near-horizon limit also suffices to capture the features of the Hawking radiation \cite{Fredenhagen:1989kr}.

\subsection{Back-reaction}

Back-reaction issues lie at the heart of any understanding of black-hole information loss, and one might wonder whether back-reaction itself changes the basic picture described above. In particular, one might wonder if the late-time breakdown of perturbation theory found here gives rise to effects that are large enough to destabilize the background geometry about which our calculations are based. 

It can (but need not) happen that late-time secular effects can mean back-reaction is large enough to significantly alter the background. The secular growth of interest here describes the accumulation over long times of perturbatively small effects that ultimately lead to significant deviations from the leading-order behaviour. Whether this also causes significant change to the background geometry depends on how much stress-energy is associated with this long buildup. 

Existing examples reveal at least two different reasons why significant back-reaction need not occur despite the late-time breakdown of naive perturbation theory. The first of these is illustrated by the Rindler example considered here. In this instance late-time perturbative breakdown occurs not because correlation functions are becoming large at late times but because they are falling, though more slowly than is the unperturbed propagator. Since large fluctuations are not present there is no reason to expect a large stress energy.

Cosmology both provides a second concrete illustration of late-time perturbative breakdown without any accompanying evidence for uncontrolled back-reaction and an example with explicitly curved backgrounds. In this case secular evolution happens because small gaussian fluctuations accumulate and resum to produce a spectrum of very nongaussian fluctuations on super-horizon scales \cite{Stochastic2, TsamisWoodard}. But when this late-time fluctuation spectrum is used to compute stress-energy expectations there is no evidence for it becoming large \cite{OpenEFT2}. 

Indeed, one of the goals of applying the above techniques to black-hole geometries is to establish precisely how back-reaction evolves at late times in a reliable way.

\medskip

In summary, our Rindler calculation suggests that perturbative methods might break down in the late-time regime for gravitational systems with horizons, that have thermal properties. This might be relevant for black-hole information puzzles, even where curvatures are weak enough that Wilsonian EFT methods would naively have been thought to be working. If true, simple extrapolations based on free fields in curved space might be misleading at very late times. The good news is that such late-time problems are likely to be resummable. We intend to pursue explicit calculations to reveal whether or not the above arguments really do undermine inferences about late-time behaviour of quantum fields near black holes.

\section*{Acknowledgements}
We thank Peter Adshead, Richard Holman and Sarah Shandera for many helpful discussions, as well as Tereza Verdanyan who contributed to early stages of this project. CB thanks the Banff International Research Station for hospitality through its `research in teams' program, during which parts of this work was developed. This work was partially supported by funds from the Natural Sciences and Engineering Research Council (NSERC) of Canada. Research at the Perimeter Institute is supported in part by the Government of Canada through NSERC and by the Province of Ontario through MRI. 

\appendix

\section{Calculation of late-time thermal behaviour}
\label{App:ThermCalc}

This appendix provides more details of the secular late-time behaviour of scalar correlation functions at finite temperature. Because our interest is in time-dependence of correlation functions we use Takahashi and Umezawa's Thermo-field dynamics \cite{Takahasi:1974zn}, a variant of the Keldysh real-time finite-temperature formalism \cite{Schwinger:1960qe, Keldysh:1964ud} rather than the Euclidean Matsubara formalism (for textbook discussions, see \cite{TFTBook1, TFTBook2, TFTBook3, TFTBook4, TFTBook5, TFTBook6, Landsman:1986uw}. 

This involves a doubling of the degrees of freedom with $\phi_1$ representing the physical field and $\phi_2$ the thermal ghost field. The extra field is introduced so that the vacuum expectation value of any physical operator agrees with its statistical average for an ensemble in thermal equilibrium (with temperature $T$) \cite{Israel:1976ur}. We make use of the Lagrangian:
\begin{eqnarray}
L & = & - \tfrac{1}{2} \partial_{\mu}\phi_1\partial^{\mu}\phi_1 - \tfrac{1}{2} m^2 \phi^2_1 - \tfrac{\lambda}{4} \phi_1^4 + \tfrac{1}{2} \partial_{\mu}\phi_2\partial^{\mu}\phi_2 + \tfrac{1}{2} m^2 \phi^2_2 + \tfrac{\lambda}{4} \phi_2^4
\end{eqnarray}

The extra degrees of freedom introduced by $\phi_2$ correspond to the particle states on the hidden side of the horizon (ie. modes in the left Rindler wedge relative to modes in the right Rindler wedge) in the context of the Rindler story \cite{Israel:1976ur}. There are four propagators $\Delta_{jk}(p;m)$ where $j,k \in \{1,2\}$ label the field type of the external points. The momentum-space propagators are:
\begin{eqnarray}
- i \Delta_{11}(p;m) & = & \frac{-i}{-p_0^2 + |\mathbf{p}|^2 + m^2 - i \epsilon} + \frac{2 \pi \delta(-p_0^2 + |\mathbf{p}|^2 + m^2 )}{e^{|p_0|/T} - 1} \label{prop11} \\
- i \Delta_{12}(p;m) \ = \ - i \Delta_{21}(p;m) & = & \pi \mathrm{csch}\left( \tfrac{|p_0|}{2T} \right) \delta(-p_0^2 + |\mathbf{p}|^2+m^2) \label{prop12} \\
- i \Delta_{22}(p;m) & = & \frac{i}{-p_0^2 + |\mathbf{p}|^2 + m^2 + i \epsilon} + \frac{2 \pi \delta(-p_0^2 + |\mathbf{p}|^2 + m^2)}{e^{|p_0|/T} - 1}
\end{eqnarray}

In what follows we will consider the massless theory $m \to 0^{+}$. The massless tadpole correction to the (physical) 11-propagator is a sum over two diagrams; the external points are fixed to be of type 1 and the vertices must be varied over types 1 {\it and} 2 \cite{Evans:1986ws, Evans:1988ub}:
\begin{eqnarray}
G_{\mathrm{tad}}(x;y)_{11} \ & = & \ \vcenter{\hbox{\includegraphics[scale=0.13]{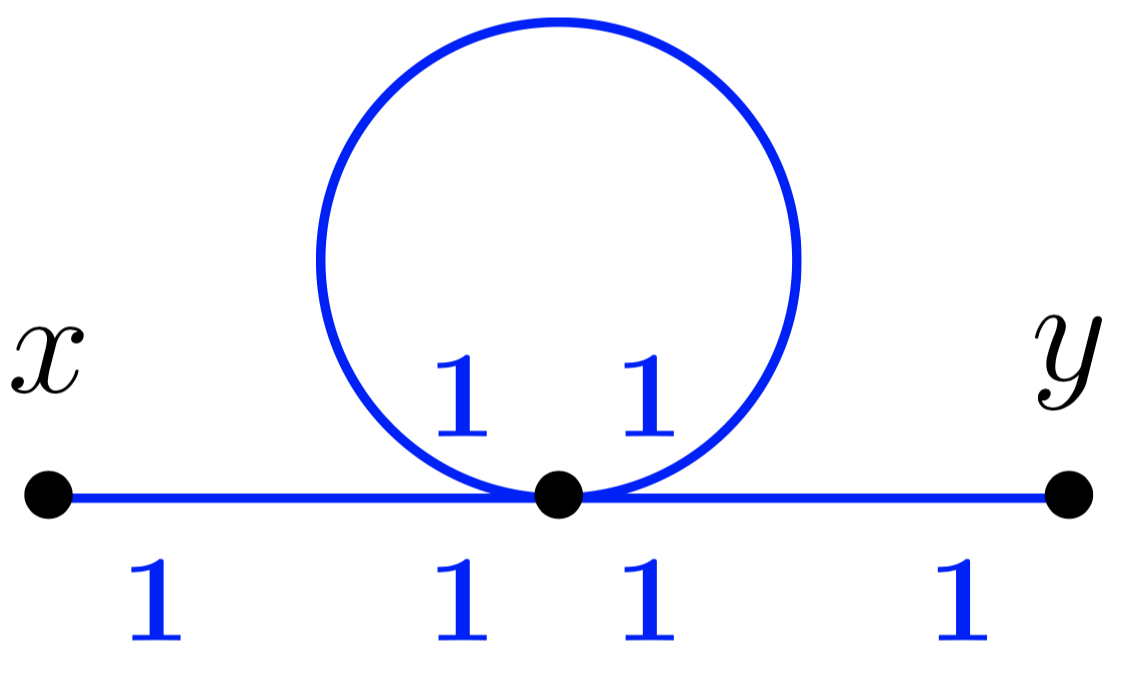}}}  \ \ \ + \ \ \ \vcenter{\hbox{\includegraphics[scale=0.13]{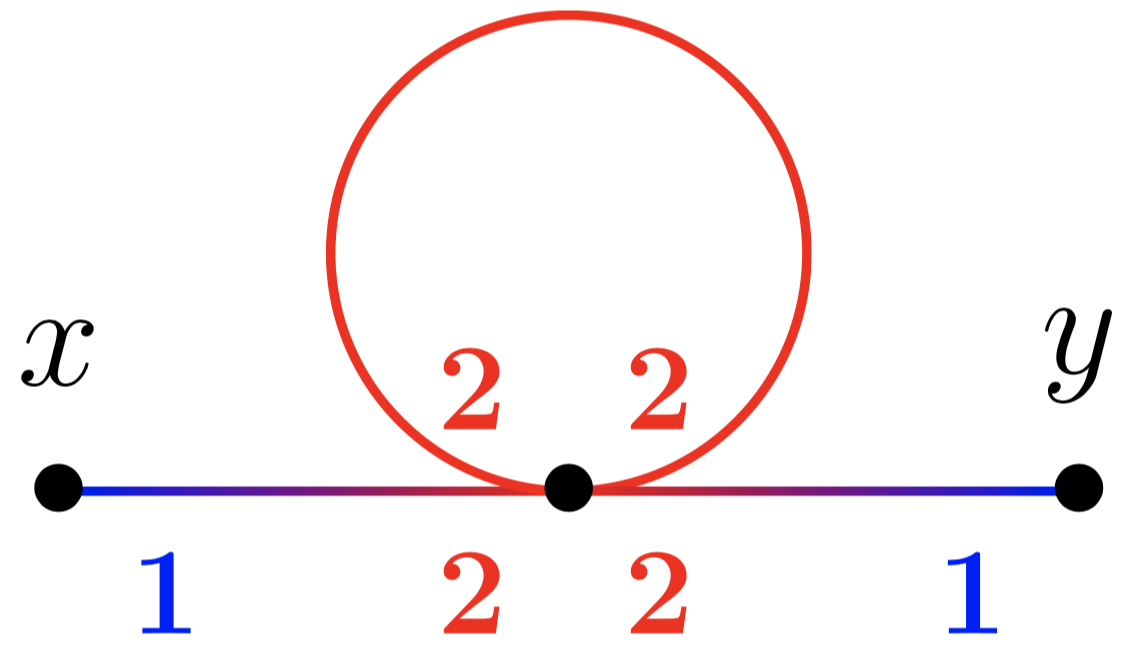}}} \\
& = &  - 3 i \lambda \left[ \int \frac{d^{4}p}{(2\pi)^4}\ \left[ - i \Delta_{11}(p;0)\right]^{2} \ e^{i p \cdot (x - y)} \right] \left[ \int \frac{d^{4}k}{(2\pi)^4}\left[ - i\Delta_{11}(k;0) \right] \right] \nonumber \\
& \ & \ \ \ \ \ \ + \ 3 i \lambda \ \left[ \int \frac{d^{4}p}{(2\pi)^4}\ \left[- i \Delta_{12}(p;0)\right]^{2} \ e^{i p \cdot (x - y)} \right] \left[ \int \frac{d^{4}k}{(2\pi)^4} \left[ - i \Delta_{22}(k;0) \right] \right] \label{ThermalTad1}
\end{eqnarray}

The loop integrals evaluate to 
\begin{eqnarray}
\left[ \int \frac{d^{4}k}{(2\pi)^4} \left[ - i \Delta_{11}(k;0) \right] \right] = \left[ \int \frac{d^{4}k}{(2\pi)^4} \left[ - i \Delta_{22}(k;0) \right] \right]^{\ast} & = & \mathcal{Z} + \mathcal{T} 
\end{eqnarray}

Where:
\begin{eqnarray}
\mathcal{Z} & = & \int_{-\infty}^{\infty} \frac{dk_0}{2\pi} \int \frac{d^{3}\mathbf{k}}{(2\pi)^3} \frac{i}{k_0^2 - |\mathbf{k}|^2 + i \epsilon} \ \to \  0 \\
\mathcal{T} & = & \int_{-\infty}^{\infty} dk_0 \int \frac{d^{3}\mathbf{k}}{(2\pi)^3} \frac{\delta(k_0^2 - |\mathbf{k}|^2)}{e^{|k_0|/T} - 1} \ = \ \frac{T^2}{2\pi^2} \int_{0}^{\infty} d\chi\ \frac{\chi}{e^{\chi} - 1} \ = \ \frac{T^2}{12}
\end{eqnarray}

The integral $\mathcal{Z}$ is the standard tadpole loop encountered in the zero-temperature theory: after dimensional regularization and the addition of an appropriate counter-term to the Lagrangian this gets renormalized to $0$. The novel thermal integral evaluates trivially. Combined with the symmetry factor $3$ and coupling $\lambda$ this accounts for the quoted thermal mass shift \cite{TFTBook4}:
\begin{eqnarray}
\delta m_{T}^2 =  \frac{\lambda T^2}{4}
\end{eqnarray}

Accounting for the above mass shift, we get the following simplification in (\ref{ThermalTad1}):
\begin{eqnarray}
G_{\mathrm{tad}}(x;y)_{11} & = & \frac{i\lambda T^2}{4} \int \frac{d^{4}p}{(2\pi)^4}\ \bigg( - \left[ - i \Delta_{11}(p;0)\right]^{2} + \left[ - i \Delta_{12}(p;0)\right]^{2} \bigg) \ e^{i p \cdot (x - y)} \label{ThermalTad2}
 \end{eqnarray}

Using the explicit forms (\ref{prop11}) and (\ref{prop12}) the function in the brackets becomes:
\begin{eqnarray}
- \big[ - i \Delta_{11}(p;0) \big]^2 + \big[ - i \Delta_{12}(p;0) \big]^2   & = & \frac{1}{\left( -p_0^2+|\mathbf{p}|^2 - i \epsilon \right)^2} + \frac{4\pi i}{e^{|p_0|/T} - 1} \frac{\delta(-p_0^2 + |\mathbf{p}|^2)}{-p_0^2+|\mathbf{p}|^2 - i \epsilon} \\
& \ & \ + \pi^2 \bigg[ \mathrm{csch}^2\left( \frac{|p_0|}{2T} \right) -  \frac{4}{(e^{|p_0|/T} - 1)^2} \bigg] \left[ \delta(-p_0^2 + |\mathbf{p}|^2) \right]^2 \nonumber
\end{eqnarray}

For any $\chi >0$ we have the identity $\mathrm{csch}^{2}(\tfrac{\chi}{2}) - \frac{4}{(e^{\chi} - 1)^2} \ = \ \frac{4}{e^{\chi} - 1}$ and so the above simplifies to:
\begin{eqnarray}
\big[ - i \Delta_{12}(p;0) \big]^2 - \big[ - i \Delta_{11}(p;0) \big]^2 & = & \frac{1}{\left( -p_0^2+|\mathbf{p}|^2 - i \epsilon \right)^2} + \frac{4\pi i}{e^{|p_0|/T} - 1} \frac{\delta(-p_0^2 + |\mathbf{p}|^2)}{-p_0^2+|\mathbf{p}|^2 - i \epsilon }  \nonumber \\
& \ & \ \ \ \ \ +  \frac{4 \pi^2  }{ e^{|p_0|/T} - 1 } \left[ \delta(-p_0^2 + |\mathbf{p}|^2) \right]^2 
\end{eqnarray}

At this point we use the regularization $\frac{\delta(z)}{z - i \epsilon} - i \pi \left[\delta(z)\right]^2 = - \frac{1}{2} \delta^{\prime}(z)$ (where $\delta^{\prime}$ is the derivative of the Dirac delta) \cite{TFTBook4} and we can write (\ref{ThermalTad2}) as:
\begin{eqnarray}
G_{\mathrm{tad}}(x;y)_{11} & = & \frac{ i \lambda T^2}{4} \ \int \frac{d^{4}p}{(2\pi)^4}\ \left[  \frac{1}{(-p_0^2 + |\mathbf{p}|^2 - i \epsilon)^2} - \frac{2 \pi i \delta'(-p_0^2 + |\mathbf{p}|^2)}{e^{|p_0|/T} - 1} \right] e^{i p \cdot (x - y)} \label{ThermalTad3}
\end{eqnarray}

For any even function $f(-\omega) = f(\omega)$ we can write its 1D Fourier transform in terms of a cosine Fourier transform as follows:
\begin{eqnarray}
\int_{-\infty}^{\infty} d\omega\ f(\omega) e^{- i t \omega} & = & 2 \int_{0}^{\infty} d\omega\ f(\omega) \cos( t \omega )
\end{eqnarray}

Furthermore for any radial function $g(\boldsymbol{\sigma}) = G(|\boldsymbol{\sigma}|)$, we can write its 3D Fourier transform as:
\begin{eqnarray}
\int \frac{d^{3}\boldsymbol{\sigma}}{(2\pi)^3} G(|\boldsymbol{\sigma}|) e^{- i \mathbf{r} \cdot \boldsymbol{\sigma}} \ = \  \frac{1}{2 \pi^2 |\mathbf{r}|} \int_{0}^{\infty} d\Sigma \ \Sigma \sin(|\mathbf{r}|\Sigma) G(\Sigma) 
\end{eqnarray}

So in terms of the dimensionless variables $\Sigma = \frac{|\mathbf{p}|}{T}$ and $\omega = \frac{p_0}{T}$ (\ref{ThermalTad3}) becomes
\begin{eqnarray}
G_{\mathrm{tad}}(x;y)_{11} & = & \frac{i\lambda T}{8\pi^3 |\mathbf{x} - \mathbf{y}|} \int_{0}^{\infty} d\omega \ \cos\left( T( x^{0}-y^0) \omega \right) \\
& \ & \ \ \ \ \times \ \int_0^{\infty} d\Sigma \ \Sigma \sin\left( T |\mathbf{x} - \mathbf{y}| \Sigma \right) \left[ \frac{1}{\left( -\omega^2+\Sigma^2 - i \epsilon \right)^2} - \frac{2\pi i\delta^{\prime}(-\omega^2 + \Sigma^2)}{e^{\omega} - 1} \right]  \nonumber
\end{eqnarray}

where we've used the scaling property $\delta'(a \chi) = \frac{\delta'(\chi)}{|a|^2}$  (immediately following from $\delta(a\chi) = \frac{\delta(\chi)}{|a|}$). Next we take the limit $\epsilon \to 0^{+}$ using the regularization $\frac{1}{\left(z - i \epsilon\right)^2} \ = \ \mathrm{Pf}\left[ \frac{1}{z^2} \right] - i \pi \delta^{\prime}(z)$ \cite{Landsman:1986uw} giving
\begin{eqnarray}
G_{\mathrm{tad}}(x;y)_{11} & = & \frac{ i \lambda T}{8 \pi^3 |\mathbf{x} - \mathbf{y}|} \ \int_{0}^{\infty} d\omega\ \cos\left(T( x^0 - y^0) \omega \right) \int_{0}^{\infty} d\Sigma\ \Sigma \sin\left(T|\mathbf{x} - \mathbf{y}| \Sigma\right) \mathrm{Pf} \left[ \frac{1}{(-\omega^2 + \Sigma^2)^2}  \right] \label{ThermalTad4} \\
& \ & \ \ \ + \frac{ \lambda T}{8 \pi^2 |\mathbf{x} - \mathbf{y}|} \ \int_{0}^{\infty} d\omega\ \cos\left( T ( x^0 - y^0 ) \omega \right) \coth\left( \frac{\omega}{2} \right) \int_{0}^{\infty} d\Sigma\ \Sigma \sin\left( T |\mathbf{x} - \mathbf{y}| \Sigma\right) \delta^{\prime}(-\omega^2 + \Sigma^2 \big) \nonumber
\end{eqnarray}

where we've made use of the identity $1 + \frac{2}{e^{\omega} - 1} = \mathrm{coth}\left( \frac{\omega}{2} \right)$. The pseudo-function $\mathrm{Pf}\left[ \frac{1}{z^2} \right]$ is the regularization of $\frac{1}{z^2}$ \cite{Landsman:1986uw} obeying $\int_{-\infty}^{\infty} dz \mathrm{Pf}\left[ \frac{1}{z^2} \right] f(z) = \mathcal{PV} \int_{-\infty}^{\infty} dz \frac{f'(z)}{z}$ (where $\mathcal{PV}$ is the Cauchy principal value) \cite{Kanwal}. We make use of a partial fraction expansion for the first $\Sigma$-integral in (\ref{ThermalTad4}) giving:
\begin{eqnarray}
\int_{0}^{\infty} d\Sigma\ \Sigma \sin\left(T|\mathbf{x} - \mathbf{y}| \Sigma\right) \mathrm{Pf} \left[ \frac{1}{(-\omega^2 + \Sigma^2)^2}  \right] & = & \frac{1}{4\omega} \int_{0}^{\infty} d\Sigma\ \sin\left(T|\mathbf{x} - \mathbf{y}| \Sigma\right) \mathrm{Pf} \left[ \frac{1}{(\Sigma - \omega)^2} \right]  \\
& \ & \ \ \ \ \  - \frac{1}{4\omega} \int_{0}^{\infty} d\Sigma\ \sin\left(T|\mathbf{x} - \mathbf{y}| \Sigma\right) \mathrm{Pf} \left[ \frac{1}{(\Sigma + \omega)^2} \right] \nonumber \\
& = & \frac{T|\mathbf{x} - \mathbf{y}|}{4\omega} \ \mathcal{PV} \int_{0}^{\infty} d\Sigma\ \cos\left(T|\mathbf{x} - \mathbf{y}| \Sigma\right) \left[ \frac{1}{\Sigma - \omega} - \frac{1}{\Sigma + \omega} \right] \ \ \ \ \ \  \ \ \  \\
& = & - \frac{\pi T|\mathbf{x} - \mathbf{y}|}{4\omega} \sin\left( T|\mathbf{x} - \mathbf{y}| \omega \right) \label{FirstSigma}
\end{eqnarray} 

Next we make note of the well-known identity $\delta(-\omega^2 + \Sigma^2) = \frac{\delta(\Sigma - \omega) + \delta(\Sigma + \omega)}{2|\omega|}$ which may be differentiated to yield \cite{Kanwal}:
\begin{eqnarray}
\delta'(-\omega^2 + \Sigma^2) = \frac{\delta'(\Sigma - \omega) + \delta'(\Sigma + \omega)}{4 |\omega| \Sigma}
\end{eqnarray}

With this and the rule $\int_{-\infty}^{\infty}dz f(z) \delta'(z) = - \int_{-\infty}^{\infty} dz f'(z) \delta(z)$ the second $\Sigma$-integral in (\ref{ThermalTad4}) may be integrated giving (recalling that $\omega>0$):
\begin{eqnarray}
\int_{0}^{\infty} d\Sigma\ \Sigma \sin\left( T |\mathbf{x} - \mathbf{y}| \Sigma\right) \delta^{\prime}(-\omega^2 + \Sigma^2 \big) & = & \frac{1}{4\omega} \int_{0}^{\infty} d\Sigma\ \sin\left( T |\mathbf{x} - \mathbf{y}| \Sigma\right) \left[ \delta^{\prime}(\Sigma - \omega \big) + \delta^{\prime}(\Sigma + \omega \big) \right] \\
& = & - \frac{T |\mathbf{x} - \mathbf{y}|}{4\omega} \int_{0}^{\infty} d\Sigma\ \cos\left( T |\mathbf{x} - \mathbf{y}| \Sigma\right) \left[ \delta(\Sigma - \omega \big) + \delta(\Sigma + \omega \big) \right] \ \ \ \ \ \ \ \ \ \\
& = & - \frac{T |\mathbf{x} - \mathbf{y}|}{4\omega} \cos\left( T |\mathbf{x} - \mathbf{y}| \omega \right) \label{SecondSigma}
\end{eqnarray}

Putting (\ref{FirstSigma}) and (\ref{SecondSigma}) all together into (\ref{ThermalTad4}) leaves us with:
\begin{eqnarray}
G_{\mathrm{tad}}(x;y)_{11} & = & - \frac{ i \lambda T^2}{32 \pi^2} \ \int_{0}^{\infty} d\omega\ \frac{\sin\left( T |\mathbf{x} - \mathbf{y}| \omega \right) \cos\left( T(x^0 - y^0) \omega \right) }{\omega}\\
& \ & \ \ \  - \frac{ \lambda T^2}{32 \pi^2 } \ \int_{0}^{\infty} d\omega\  \frac{ \coth\left( \frac{\omega}{2} \right) \cos\left( T |\mathbf{x} - \mathbf{y}| \omega \right) \cos\left( T( x^0 - y^0) \omega \right) }{\omega} \nonumber
\end{eqnarray}

And setting $x^0 - y^0 = t$ and $\mathbf{x} = \mathbf{y}$ we have:
\begin{eqnarray}
G_{\mathrm{tad}}(x;y)_{11} & = & - \frac{ \lambda T^2}{32 \pi^2} \ \int_{0}^{\infty} d\omega\  \frac{ \coth\left( \frac{\omega}{2} \right) }{\omega} \cos\left( T t \omega \right)
\end{eqnarray}

We interpret the singular distribution being cosine Fourier transformed in the following manner:
\begin{eqnarray}
\frac{ \coth\left( \frac{\omega}{2} \right) }{\omega} = \frac{2}{\omega^2} + \left[ \frac{ \coth\left( \frac{\omega}{2} \right) }{\omega} - \frac{2}{\omega^2} \right]
\end{eqnarray}

This distribution has one singularity of the form $\frac{2}{\omega^2}$ and the remainder function $\frac{ \coth\left( \frac{\omega}{2} \right) }{\omega} - \frac{2}{\omega^2}$ has absolutely integrable $N^{\mathrm{th}}$ derivatives for all $N \geq 1$ on the real line, and also falls to zero at $\omega \to \infty$. This means that the asymptotic form of the function in the limit $Tt \to \infty$ is governed by the Fourier cosine transform of $\frac{2}{\omega^{2}}$ which is $- \pi \big|T t\big|$ \cite{Lighthill} giving the result (\ref{Tsect}):
\begin{eqnarray}
G_{\mathrm{tad}}(x;y)_{11} & = & \frac{1}{32 \pi} \lambda T^3 |t| \ + \ \mathrm{subdominant}
\end{eqnarray}

Or in terms of the thermal mass shift $\delta m_{T}^2 =  \frac{\lambda T^2}{4}$ we get that in the limit $Tt \to \infty$:
\begin{eqnarray}
G_{\mathrm{tad}}(x;y)_{11} & = & \frac{\delta m_{T}^2 }{8 \pi} T |t| \ + \ \mathrm{subdominant}
\end{eqnarray}

This is in agreement with (\ref{GRind1}).

\end{document}